\documentclass{iaus}
\usepackage{graphicx}
\pdfoutput=1

\title[Getting ready for the micro-arcsecond era]{Getting ready for the micro-arcsecond era}

\author[Anthony G.A.\ Brown]{Anthony G.A.\ Brown}

\affiliation{Sterrewacht Leiden, Leiden University, \\ P.O.\ Box 9513, 2300 RA
Leiden, The Netherlands \\ email: \texttt{brown@strw.leidenuniv.nl}}

\pubyear{2008}
\volume{248}
\pagerange{XXX--YYY}
\jname{A Giant Step: from Milli- to Micro-arcsecond Astrometry}
\editors{W.J. Jin, I. Platais \& M.A.C. Perryman, eds.}

\begin{document}

\maketitle

\begin{abstract}
  As the title of this symposium implies, one of the aims is to examine the
  future of astrometry as we move from an era in which thanks to the Hipparcos
  Catalogue everyone has become familiar with milliarcsecond astrometry to an
  era in which microarcsecond astrometry will become the norm. I will take this
  look into the future by first providing an overview of present astrometric
  programmes and how they fit together and then I will attempt to identify the
  most promising future directions. In addition I discuss the important
  conditions for the maximization of the scientific return of future large and
  highly accurate astrometric catalogues; catalogue access and analysis tools,
  the availability of sufficient auxiliary data and theoretical knowledge, and
  the education of the future generation of astrometrists.
  \keywords{astrometry, catalogs, surveys}
\end{abstract}

\firstsection

\section{Overview of astrometric programmes}

\begin{figure}[ht]
  \includegraphics[width=\textwidth]{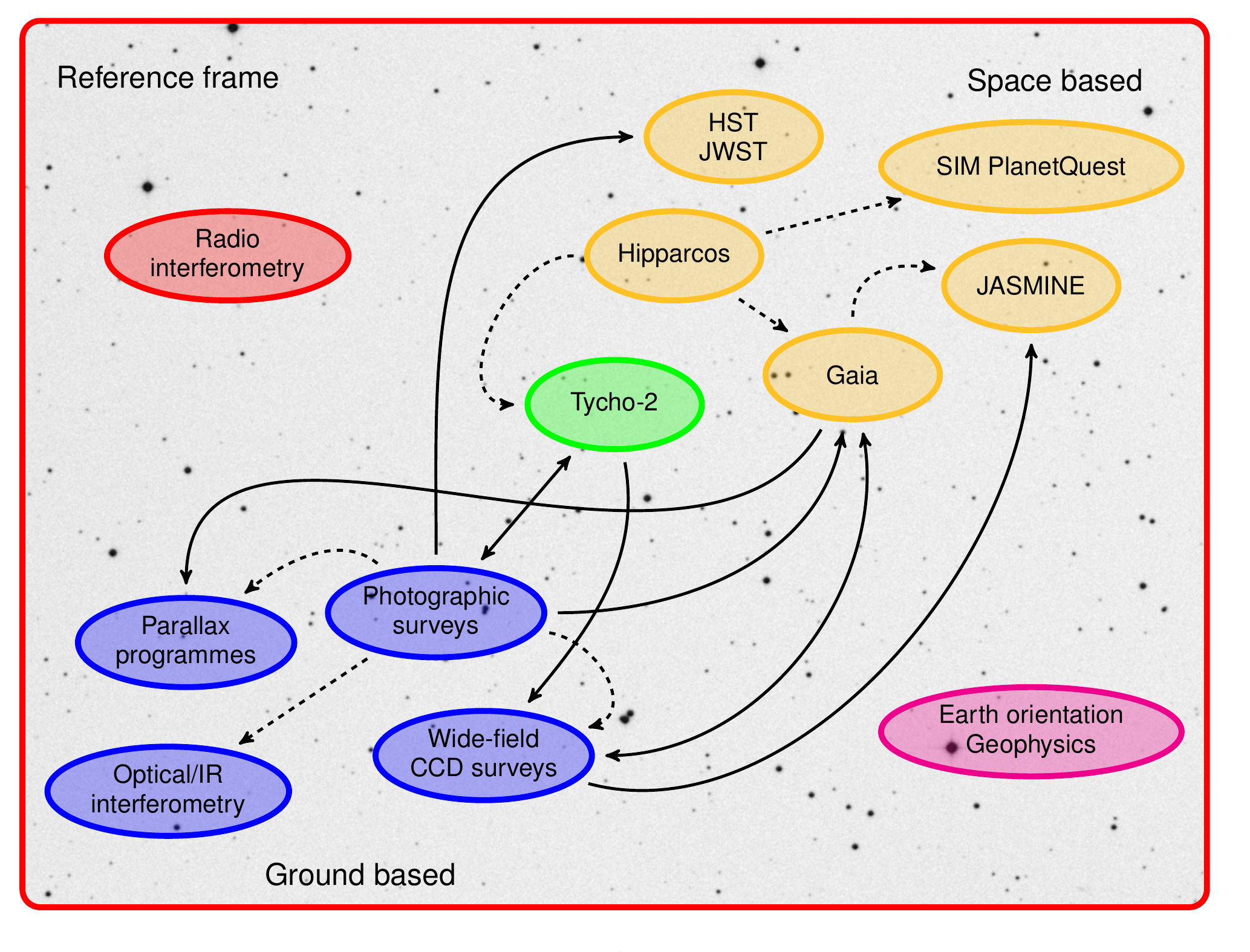}
  \caption{Graphical overview of astrometric programmes and the relations
  between them. There is no ordering in terms of timing or accuracy of the
  programmes. Refer to the text for further explanations of this figure. The
  background image, symbolizing the reference frame, is a $20\times 15$
  arcmin$^2$ region of the sky centred on the Hyades cluster and was taken from
  the Digitized Sky Survey (POSS-II plates).}
  \label{fig:overview}
\end{figure}

Figure~\ref{fig:overview} presents a graphical overview of past, present, and
future astrometric programmes and the relationships between them. I have grouped
a number of the programmes together. For example, `wide-field CCD surveys'
stands for astrometry obtained from observing programmes such as the Sloan
Digital Sky Survey (SDSS) or the future Large Synoptic Survey Telescope
\cite[(LSST, see Ivezi\'c 2008)]{Ivezic2008}. Similarly `radio interferometry'
stands for a number of astrometric programmes conducted with different VLBI
networks. The overview in figure~\ref{fig:overview} is not intended to be
complete but serves to illustrate the variety of astrometric programmes and how
they fit together.

All astrometric programmes must ultimately be linked to the International
Celestial Reference System (ICRS, the idealized barycentric coordinate system to
which celestial position are referred) through procedures which involve a number of
steps that depend on the details of the astrometric observing programme. For
ground-based programmes where the observations are typically done over a small
field of view the obtained relative parallaxes and proper motions are converted
to an absolute scale through the observations of astrometric standards. In the
case of missions like Hipparcos although the measured parallaxes and proper
motions are absolute by design, the reference frame still has six degrees of
freedom with respect to the ICRS (three orientation and three spin parameters)
which have to be taken out through observations of extra-galactic reference
sources. The ICRS thus ties all astrometric programmes together and this is
symbolized by the background image in figure~\ref{fig:overview}.

The practical realization of the ICRS is the International Celestial Reference
Frame (ICRF) consisting of a few hundred extra-galactic radio sources with
adopted ICRS positions. The ICRF sources are constantly monitored in dedicated
VLBI observing programmes, which is why the radio interferometry observations are
portrayed separately in figure~\ref{fig:overview} as they underpin the practical
realization of the reference system. In the optical, the ICRF is materialized by
the Hipparcos Catalogue. For more details see the chapters on reference frames
in this volume. For Solar system investigations radar astrometric observations
(ranging and Doppler) are instrumental in providing the reference for accurate
ephemerides of the inner planets and the determination of various astronomical
constants (see Pitjeva in this volume).

A well established reference system can be used to interpret accurate
observations of celestial sources from earth in terms of the orientation of the
earth's spin axis and its precession and nutation (bottom right of
figure~\ref{fig:overview}). Two particularly interesting examples are given in
the contributions by Vondr\'ak and Huang in this volume. The former concerns a
study of earth rotation over the past 100 years and the latter a study of the
precession and nutation of the earth's rotation axis in terms of a non-rigid
model of our planet.

The other astrometric programmes in figure~\ref{fig:overview} have been lumped
together into the ca\-tegories `ground based' and `space based', with the
Tycho-2 catalogue located in between as that was constructed from a combination
of photographic plate material and observations made with the Hipparcos
satellite. The arrows illustrate some of the relations within the ground based
or space based programmes (dashed lines) and between ground based and space
based astrometry (solid lines).

The photographic surveys that were carried out during the 20th century have
formed the basis for the optical reference frame until the Hipparcos catalogue
was published and the astrometric catalogues derived from them, now reduced to
the Hipparcos reference frame, still serve as basic input for modern ground
based astrometric programmes. Recently the photographic material has been
combined with CCD astrometric observations such as in the UCAC catalogues
\cite[(Zacharias et al.\ 2004)]{Zacharias2004}.  Eventually the wide field CCD
surveys will provide a much denser optical reference frame, although they will
still have to be linked to more absolute frames, e.g.\ Hipparcos or Gaia.

A good example of how ground based astrometric surveys support the space
missions is the Guide Star Catalogue \cite[(GSC, Bucciarelli
2008)]{Bucciarelli2008}.  It serves as the basic input for HST pointing and
guidance and will also form the basis of the input catalogue for JWST. For Gaia
the GSC will be one of the inputs for the construction of an initial source list
to support the identification of sources in the early phases of the mission. The
latest version of the GSC makes use of Tycho-2 data as a reference for the
astrometric calibration and as a supplement to its bright end.

The ground based CCD surveys go much deeper than missions like Gaia or JASMINE
will and the corresponding astrometric catalogues can serve to provide
information on sources beyond the survey limits of these missions. This is
important in regions on the sky where the density is high enough for faint
sources to disturb the astrometry and photometry of the target sources.
Conversely, the ground based CCD surveys can make use of the reference frame
provided by space based missions. The Tycho-2 catalogue is already being used
for this purpose and in the future the Gaia and SIM PlanetQuest catalogues will
provide an excellent reference, in particular for the ground based parallax
programmes \cite[(Smart 2008)]{Smart2008}.

Finally, the space astrometry missions are also interrelated. The Hipparcos
catalogue has been used to select candidate grid stars for SIM PlanetQuest
\cite[(see e.g.\ Hekker et al.\ 2006)]{Hekker2006} and will be used to predict
the positions of sources on the sky that are too bright to be observed by Gaia.
The latter will affect the measurements of nearby fainter sources and will
influence the state of Gaia's CCDs with respect to charge transfer inefficiency
\cite[(see Lindegren 2008)]{Lindegren2008}. The JASMINE mission will rely on a
plate overlap technique and observe only a relatively small part of the sky.
Thus this mission will benefit greatly from the reference frame which will be
provided by Gaia.

In summary all the astrometric programmes schematically indicated in
figure~\ref{fig:overview} form a very tightly interlocked set of observational
programmes, none of which be obsolete for some time to come. In the next section
I will look at how these astrometric programmes cover the space of survey
parameters and what science is consequently not well covered.

\section{Science covered by astrometric surveys: are we missing something?}

\begin{figure}[ht]
  \includegraphics[width=\textwidth]{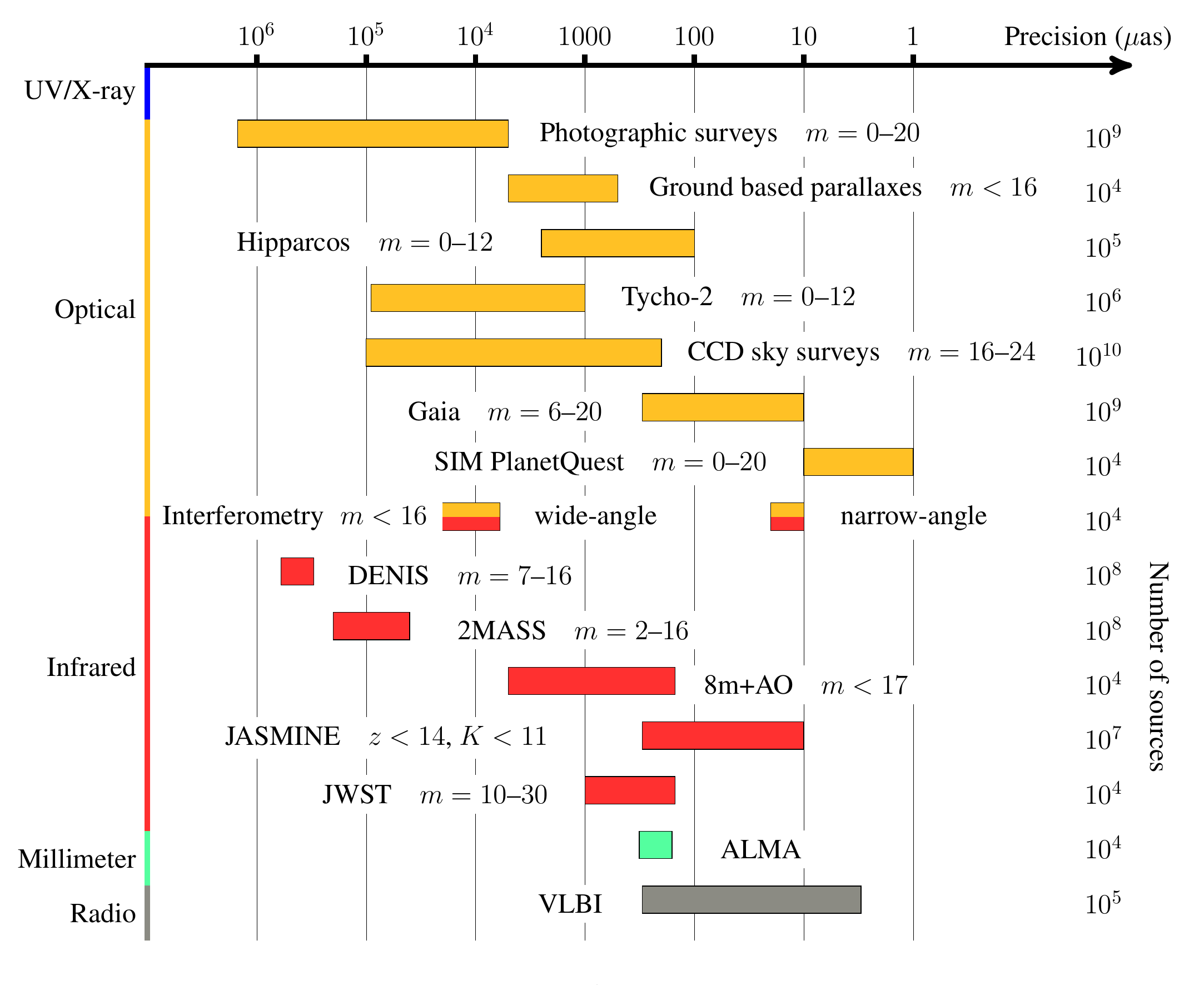}
  \caption{Graphical overview of astrometric surveys. The shading of the boxes
  and the right vertical axis serves to aid the eye in segregating the
  wavelength domains. Refer to the text for further explanation of the diagram.
  The survey parameters are based on the following references: Photographic
  surveys --- \cite[Morrison et al.\ (2001)]{Morrison2001}, \cite[Monet et al.\
  (2003)]{Monet2003}; Ground based parallaxes --- \cite[Jao (2008)]{Jao2008},
  \cite[Smart (2008)]{Smart2008}; Hipparcos --- \cite[ESA (1997)]{HIP},
  \cite[van Leeuwen (2007)]{FvL2007}; Tycho-2 --- \cite[H{\o}g et al.\
  (2000)]{Tycho-2}; CCD sky surveys --- \cite[Ivezi\'c (2008)]{Ivezic2008},
  \cite[Magnier (2008)]{Magnier2008}; Gaia --- \cite[Lindegren
  (2008)]{Lindegren2008}; SIM PlanetQuest --- \cite[Shao (2008)]{Shao2008};
  Optical/IR interferometry --- \cite[Boden (2008)]{Boden2008}, \cite[Richichi
  (2008)]{Richichi2008}; DENIS --- \cite[Borsenberger et al.\ (2006)]{DENIS},
  G.\ Simon priv.\ comm.; 2MASS --- \cite[Skrutskie et al.\ (2006)]{2MASS};
  8m+AO \cite[R\"oll et al.\ (2008)]{Roll2008}, \cite[Ghez (2008)]{Ghez2008};
  JASMINE --- \cite[Gouda (2008)]{JASMINE}; JWST --- \cite[Beichman
  (2008)]{JWST}; ALMA --- \cite[Lestrade (2008)]{Lestrade2008}; VLBI ---
  \cite[Reid (2008)]{Reid2008}, \cite[Ma (2008)]{Ma2008}.}
  \label{fig:surveys}
\end{figure}

Astrometric programmes can be characterized by the following parameters:

\begin{description}
  \item[Precision and accuracy: ] These are the most obvious selling points of
    any astrometric survey and much effort goes into maximizing these
    parameters. The precision reflects the quality of the relative measurements
    that can be made with a particular instrument and optimizing this parameter
    is enough if one is only interested in differential measurements, such as
    when hunting for exoplanets. The accuracy of a survey is limited by the
    external calibration of the measurements and by the design of the
    experiment. A well known example is the case of ground-based parallax
    programmes which are done over small fields of view and thus require
    corrections of precisely measured relative parallaxes to absolute ones. The
    accuracy in this case is limited by our knowledge of the spatial
    distribution of the reference sources.

  \item[Time baseline: ] Parallax and proper motion measurements benefit from a
    long time baseline, which is needed to disentangle the parallaxes from the
    proper motions and to ensure that proper motions of binary systems reflect
    the centre of mass motion as opposed to the orbital motions of the
    components. The sampling in time also plays an important role, especially
    for parallax measurements carried out over only a few years.

  \item[Depth and dynamic range: ] The faint limit of the survey determines the
    kinds of sources that can be observed (faint nearby low-mass objects,
    distant standard candles, extra-galactic reference sources, etc). Future
    astrometric surveys will push this parameter as much as possible but it has
    to be traded off against the dynamic range of the survey. Instruments
    capable of observing very faint sources often cannot deal with very bright
    sources. For example the Hipparcos astrometry for the brightest ($V<6$)
    stars will not be superseded by any of the planned space missions.  On the
    other hand optical/infrared interferometry programmes can typically only
    deal with bright sources.

  \item[Number of sources: ] For sources of arbitrary brightness achieving high
    precision always comes at the expense of more observing time and so a
    trade-off has to be made between the number of sources observed and the
    astrometric precision achieved. Targeted high precision programmes (such as
    SIM or the ground-based parallaxes) are very good at addressing specific
    scientific questions whereas less precise `all-sky' surveys are
    scientifically important because of the advantages of statistical studies of
    large samples and they come with the added bonus of serendipity.

  \item[Density of sources: ] The density of sources is important if the
    astrometric catalogue is to be used as a reference frame for other
    astrometric programmes. This is especially needed for astrometric surveys
    targeting faint sources over small fields of view.

  \item[Wavelength range: ] This parameter sets the `window on the universe'
    for an astrometric survey and is a very important driver for the science
    that one can do.

  \item[Narrow- vs.\ wide-angle: ] In ground-based astrometry programmes
    sub-milliarcsec to micro-arcsec precision can be achieved but only through
    differential measurements over small fields. This makes it difficult to
    construct reference frames free from zonal systematic errors and
    necessitates the correction from relative to absolute parallaxes. On the
    contrary astrometry over wide angles by design gives absolute parallaxes and
    a rigid reference frame \cite[(e.g.\ Lindegren 2005)]{Lindegren2005}. The
    problem of converting relative to absolute astrometry can in principle be
    overcome for narrow angle measurements if the observations go deep enough to
    find enough suitable extra-galactic reference sources in the field of view.

  \item[Ground based vs.\ space based: ] This parameter is closely related to
    the previous one. The choice of going to space offers the advantages of a
    stable thermal environment, freedom from gravity and the atmosphere, and
    full-sky visibility \cite[(Lindegren 2005)]{Lindegren2005}. These factors in
    fact enable wide-angle astrometry as implemented on missions such as
    Hipparcos, Gaia, and SIM PlanetQuest.

\end{description}

Figure~\ref{fig:surveys} is my attempt at summarizing the coverage of the
parameter space described above by the present and planned astrometric
catalogues and surveys. The main goal of this diagram is to identify major gaps
in the parameter space and not to provide a complete and precise overview of all
astrometric surveys or programmes. Figure~\ref{fig:surveys} shows the
distribution of astrometric programmes over wavelength, precision, depth and
dynamic range, and the number of sources observed. The precision figures
encompass positions, parallaxes and proper motions and in most cases also
reflect the accuracy of the survey. Precision numbers for proper motion are
generally numerically smaller than those for positions, especially for
photographic surveys. The magnitude limits and numbers of objects are crude
indicators. Within each wavelength range the surveys have been ordered roughly
in time. From figure~\ref{fig:surveys} a number of observations can be extracted
which can help in identifying interesting future directions for astrometry.

In the optical we can expect an enormous amount of astrometric data, with
accuracies in the $10$~microarcsec to $10$~milliarcsec range, coming in over the
next 10 to 15 years from the CCD surveys of the sky (LSST, Pan-Starrs) and Gaia.
A preview of what we can expect from this data is given in the contribution of
Ivezi\'c in this volume in which he discusses the power of using SDSS Data
Release 6 for studies of the structure of the Milky Way. The latest SDSS data
release also contains proper motions with $\sim3$--$6$ mas/yr errors out to
$r=20$ for $30\times10^6$ stars \cite[(Munn et al.\ 2004)]{Munn2004}. This
offers photometry for a huge number of stars over a large volume in our galaxy
combined with kinematic data at a level of accuracy which enables both a
quantitative and qualitative breakthrough. This dataset will be dwarfed by the
combined Gaia/LSST/Pan-Starrs data which will constitute Hipparcos quality
astrometry over the magnitude range $20$--$24$!

The ground-based parallax programmes and the optical/IR interferometric
observations will fill very important niches at the bright end in the
characterization of the Sun's nearest neighbours and in the hunt for exoplanets.

From figure~\ref{fig:surveys} it is very clear that the SIM PlanetQuest mission
is unique in its capability of providing astrometry in the $1$--$10$~$\mu$as
range all the way out to 20th magnitude. It will thus be the only instrument
capable of directly addressing detailed questions about the outer reaches of our
galaxy or nearby external galaxies.

All the optical programmes obviously suffer from the extinction barrier due to
dust, especially in the Milky Way disk and towards the Galactic centre. Infrared
surveys are thus essential in achieving a complete understanding of our galaxy
and for studying star formation in all its aspects. In the infrared the
astrometric data is currently of relatively poor quality compared to what is
available in the optical even though the DENIS and 2MASS surveys represent a
tremendous step forward compared to the situation 10 years ago. The only ongoing
astrometric programmes in the infrared are the interferometric and adaptive
optics programmes which mainly deliver relative astrometry over small fields of
view (although at high precision). The only planned future infrared astrometric
survey is the JASMINE mission which will cover the Bulge and inner Milky Way
disk regions.

A very exciting prospect in the near future is the possibility of obtaining
milliarcsecond astrometry in the millimetre wavelength regime. Combined with the
already available radio astrometry at this accuracy this will enable the direct
determination of the distances and kinematics of deeply embedded star clusters
or even proto-stellar clusters (see contributions by Loinard and Lestrade).

\section{Future directions}

From the discussion of parameter space coverage by current and planned
astrometric programmes a number of gaps can be identified from which proposals
for future astrometric observing programmes can be extracted. I briefly discuss
here a number of possibilities, some of which are already covered by planned
instruments.

\begin{description}
  \item[All-sky infrared astrometry at high accuracy: ] JASMINE covers the
    inner Milky Way disk only and does not go very deep. I think a deep all sky
    astrometric survey in the infrared provides the biggest `discovery space',
    offering a tremendous improvement in our knowledge of the Milky Way disk,
    spiral arms and star forming regions. In addition low mass stars, brown
    dwarfs and `free-floating planets' can be studied over a much larger volume.
    Such a mission is probably most efficiently carried out as a scanning
    spacecraft using the principles of Hipparcos and Gaia. The question is
    whether it will be technically feasible for a reasonable cost. Can a large
    cryogenic instrument be realised which can operate over the length of time
    required for high precision astrometry? The experience from the design of
    missions like ESA's DARWIN will be very useful in this respect.

  \item[High-precision narrow angle astrometry: ] This is an area which is
    already under development through the interferometric and adaptive optics
    programmes at the current generation of large (8~m class) telescopes.
    Pushing the precision of small fields of view as much as possible towards
    the $\mu$as mark will increase the sensitivity of exoplanet searches to the
    level where earth-sized planets around nearby stars can be discovered; allow
    us to probe space-time near the massive black hole in the centre of the
    Milky Way; and enable the study of stellar orbits around the central black
    holes in external galaxies. This increase in astrometric accuracy can be
    achieved with new instruments on 8~m class telescopes used as
    interferometers and with instruments for the future extremely large
    telescopes (see the contributions by Ghez and Richichi in this volume).

  \item[Geometric distances and motions of galaxies in the local volume: ] The
    study of the dynamics of galaxies in the local volume is currently limited
    by the lack of knowledge of their proper motions. As discussed in the
    contribution by Bruntahler proper motions can be obtained as far out as the
    M31 subgroup through VLBI measurements of masers with respect to background
    quasars. He also discusses the possibility to go beyond the local group. An
    important goal for the future is to extend these studies by improving the
    precision and the enlarging the sample.

  \item[Micro-arcsecond astrometry to 25th magnitude: ] An obvious future
    instrument to consider would be a `super-Gaia', capable of delivering
    microarcsecond astrometry to 25th magnitude. This would require careful
    consideration of a number of issues. What will be gained scientifically? Is
    Hipparcos quality data to 24th magnitude enough? The answer to this question
    will depend on what we learn from Gaia and the CCD surveys. The technical
    difficulties of simply scaling up Gaia will be driven by the need to collect
    2 to 4 orders of magnitude more photons over the magnitude range 20--25.
    This cannot be achieved through more sensitive detectors (the Gaia CCDs
    already having $\sim80$\% quantum efficiency) but will have to come from
    larger optics and a larger field of view. This will make the requirements on
    instrument stability very difficult to achieve (thermal and mechanical
    design) if not impossible. It is probably better to think about alternative
    ways of efficiently doing all-sky absolute astrometry at the microarcsecond
    level.

  \item[Astrometry at wavelengths below 300~nm: ] I included the UV/X-ray
    wavelength regime in figure~\ref{fig:surveys} to highlight that currently
    there is no high precision astrometry available shortward of optical
    wavelengths. It is not clear that efforts to acquire high accuracy
    astrometry in UV/X-ray domain will pay off. It is technically difficult and
    most UV/X-ray sources may have optical counterparts which can be measured
    astrometrically.

\end{description}

\section{Maximizing the scientific return}

Although thinking about future directions for astrometric surveys is interesting
it is much more important to ensure that we optimize the scientific exploitation
of the large amounts of high accuracy astrometric data we can expect over the
next two decades. This concerns the following issues:

\textbf{Complementary data:} All the large surveys (Gaia, LSST, Pan-Starrs) will
deliver multi-colour photometry which is essential in the scientific
exploitation of the astrometric data. Photometry provides the astrophysical
characterization of the stars observed and allows the derivation of photometric
distances. The latter can be calibrated using Gaia data for bright stars and
will provide more reliable distances for the faint stars than can be derived
from parallaxes. In the case of Gaia, at the faint end of the survey the low
resolution prism spectra will not have much discriminating power with respect to
surface gravity and metallicity of the stars. This can be remedied (for most of
the sky) by supplementing Gaia photometry with the data from LSST and
Pan-Starrs. Hence collaboration between these projects is essential.

Phase space for stars and galaxies consists of six dimensions but in the surveys
targeting the stars in our galaxy the 6th dimension, radial velocity, will not
be measured for the majority of the $10^{10}$ stars for which astrometric data
will be collected. The RAVE project \cite[(Steinmetz et al.\ 2006)]{RAVE} will
collect radial velocities for $\sim10^6$ stars and Gaia will do so for
$\sim10^8$ stars. The rewards for gathering radial velocities for the remaining
stars are substantial and we should think about efficient ways of doing so. The
accuracies of the radial velocities should be matched to the astrometric
accuracies (translated to linear velocities) which for the faint end of the
surveys leads to rather modest requirements and which can be met by low
resolution spectrographs optimized for faint stars. From the ground a radial
velocity survey, even with large multi-object instruments, may be very difficult
to complete for the whole sky to 20--24th magnitude and dedicated follow-up
programmes to pursue specific questions appear more viable. From space an
all-sky survey may be possible using a scanning satellite optimized for radial
velocities.

An important issue concerning radial velocities is the accuracy of the
zero-point of the velocity scale. Given the huge efforts to acquire highly
accurate absolute astrometry, the radial velocities should be absolute to the
same level of accuracy. This point has not been addressed over the last decade
as most of the effort has gone into refining the relative precision of radial
velocity surveys in order to optimize them for planet detection. A dedicated
effort is needed now to improve the zero-point accuracy across the
Hertzsprung-Russell diagram. Astrometric data from Gaia can be used for this
through the astrometric radial velocity technique \cite[(Dravins, Lindegren, \&
Madsen 1999)]{Dravins1999} which provides radial velocities free of effects
intrinsic to the star.

The compromise between resolution and depth of the spectroscopic surveys carried
out by the RAVE project and Gaia is driven by the need for high resolution
spectroscopy for the detailed determination of the astrophysical parameters of
stars, including rotational velocities. Gathering complementary high resolution
spectra is a very expensive undertaking and is probably best done with well
designed follow-up projects that serve, for example, to better calibrate
photometric indicators or answer specific questions.

The last remark on complementary data concerns the determination of accurate
reddening laws and extinction towards the stars and galaxies observed.
Especially for the subset of stars with very high accuracy parallaxes the
knowledge of reddening and extinction will become a limiting factor in the
determination of luminosities and stellar parameters and hence in the
improvement of our understanding of stellar structure and evolution. There is a
dedicated effort within the Gaia data processing consortium to ensure a proper
determination of reddening and extinction. This effort will benefit greatly from
collaboration with other survey projects and from investigating the use of
complementary data such as infrared photometry and direct measurements of the
gas and dust components of the Milky Way.

\textbf{Preparing for the analysis of large astrometric/photometric catalogues:}
Optimizing the science return depends critically on having analysis tools and
theoretical models that enable a proper interpretation of the vast quantities of
high accuracy astrometric and photometric data. For stars the determination of
the fundamental parameters luminosity, age, mass, and chemical composition
depends heavily on the quality of stellar structure and atmosphere models. As
discussed in the contributions by Lebreton and Chaboyer many uncertainties still
remain in stellar models. Although addressing these problems will depend on the
availability of highly accurate astrometric data, steps can already be taken now
to improve our understanding of stars through better theoretical models, the
development of 3D model atmospheres and laboratory experiments aimed at
obtaining better opacities, thermonuclear reaction rates and equations of state.

For the interpretation of the astrometric data in terms of Galactic structure
and dynamics analysis tools should be developed that can deal with data for
$10^9$ stars. Specific consideration should be given to the way parallax data
is treated in order to avoid the biases that can be introduced when converting
relatively imprecise parallaxes to distances. The best way forward may be to
design analysis methods that rely on predicting the observations of Galactic
phase space rather than the phase space variables themselves.

\begin{figure}[ht]
  \includegraphics[width=0.45\textwidth]{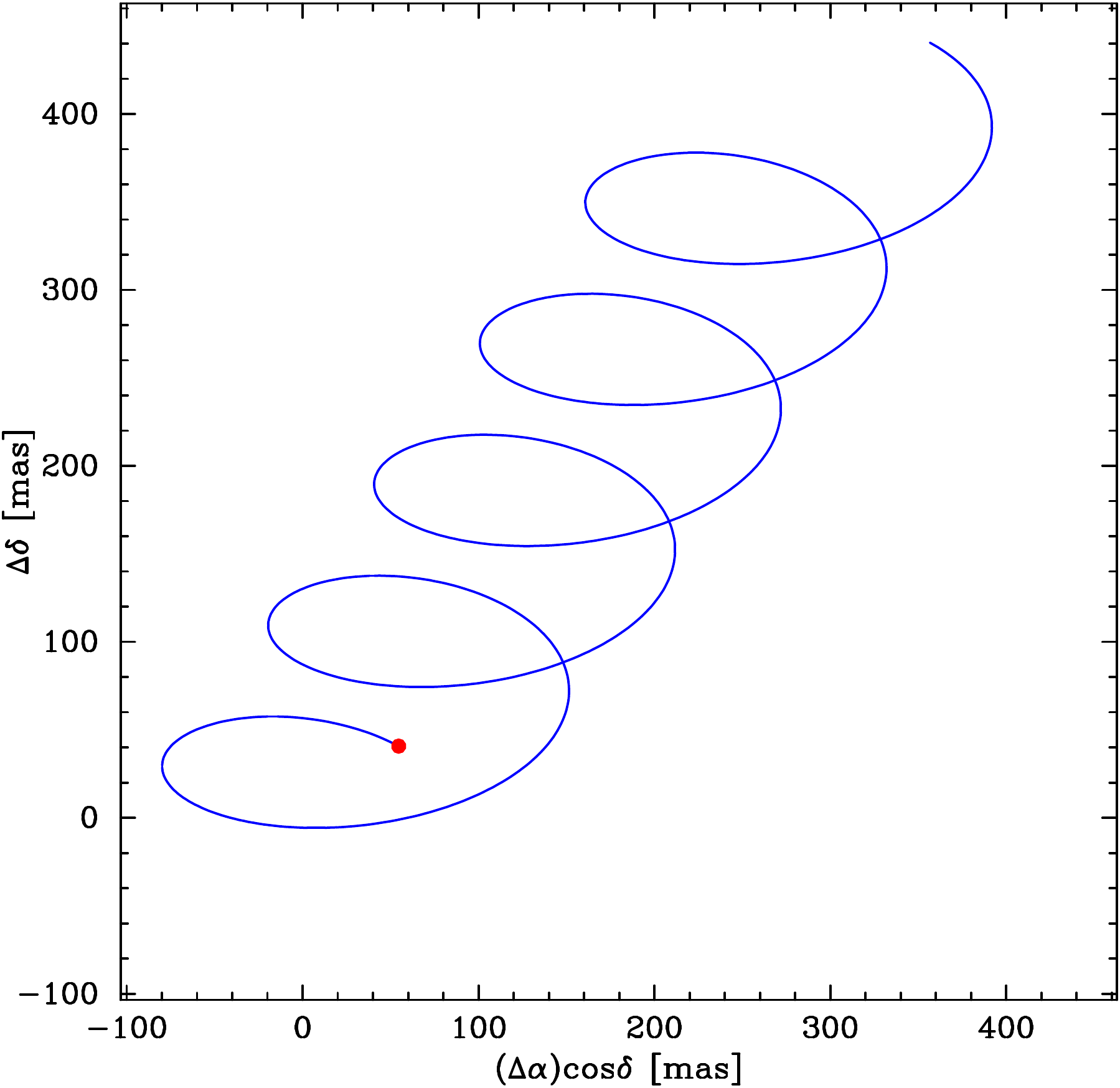}\hfill
  \includegraphics[width=0.45\textwidth]{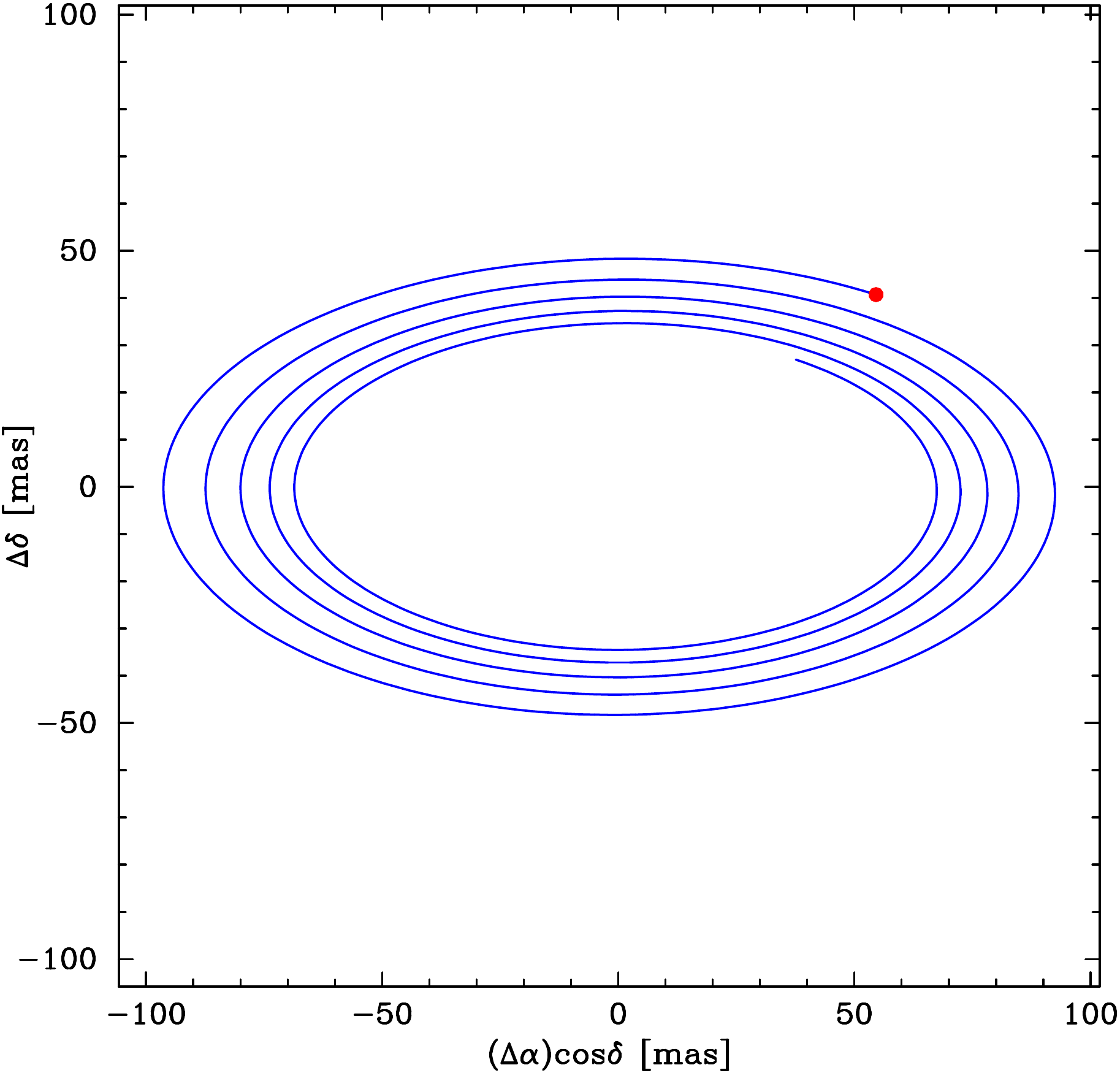}
  \caption{The paths on the sky for a star with a given proper motion and
  parallax (left) and for a star moving radially away from the Sun (right). This
  illustrates that from astrometry alone the 3D positions and motions of stars
  (including radial velocities) can be inferred. The right hand figure is not
  relevant in practice but as discussed in the paper by \cite[Dravins et al.\
  (1999)]{Dravins1999} there are a number of other astrometric effects from
  which radial velocities can be inferred. A practical application using
  Hipparcos data is given in \cite[Madsen et al.\ (2002)]{Madsen2002}. This
  figure was produced with the aid of the IAU's SOFA library
  (http://www.iau-sofa.rl.ac.uk/).}
  \label{fig:skypaths}
\end{figure}

\textbf{Providing optimized access to large catalogues:} If the data for
$10^9$--$10^{10}$ stars were to land on one's desk today it would be very
difficult to decide how to begin exploring these data and extracting interesting
science. The accessibility of these large catalogues will be essential in
determining their success. The catalogue producers should aim at making tools
available that allow easy access trough both high-level, easy to use, and
low-level, more complex, interfaces. This includes good visualization tools. The
catalogues should not be static finished products but should allow users to
re-process data based on newly acquired knowledge about certain objects or
classes of objects (see for example the discussion by Pourbaix in this volume).
Improved knowledge should be added to these databases when it becomes available.
This could be, for example, radial velocities from follow-up programmes or a
better astrophysical parametrization of stars based on improvements in
stellar atmosphere models.

Users of the catalogues should be educated through documentation which has to be
designed to be easy to understand for beginners while at the same time allowing
experts to quickly navigate to the relevant details.

\textbf{Training the next generation of astrometrists:} The design and operation
of astrometric programmes is a highly specialized undertaking and it is
essential to ensure continuity of the available expertise in the astronomical
community. This expertise should cover all aspects of carrying out astrometric
programmes, be they scanning or interferometric space missions or ground based
astrometry programmes, ranging from the establishment of the reference frame
with VLBI observations to `classical' parallax programmes. In addition
astrometry at the microarcsecond level requires orders of magnitude better
control over instrument calibration and the routine inclusion of effects such as
general relativity in the data analysis. Without dedicated astrometry experts
the whole edifice illustrated in figure~\ref{fig:overview} will fall apart. The
experts are also needed to ensure a thorough understanding of the contents and
the interpretation of astrometric catalogues. At this symposium a session was
therefore dedicated to astrometry education, and I refer to the corresponding
chapter in this volume for a summary.

\section{Final remarks}

I will conclude with a little `pep-talk'. Astrometry is unfortunately
considered to be a rather unglamourous discipline which is often even put aside as
not being `science'. I disagree with this; designing and carrying out high
accuracy astrometric experiments is a highly scientific undertaking in itself
and the results are of fundamental importance to astronomy, astrophysics and
geodynamics. Astrometry tells us about our place in the universe, provides the
only direct distance measurements to the stars, and can even be used to look
inside our own planet as discussed by Huang in this volume. If we seek to
motivate people to pursue a career in astrometry we should advertise that the
beauty of this discipline, illustrated in figure~\ref{fig:skypaths}, lies in
that fact that:

\begin{quote}
  \textit{By carefully watching the positions of stars on the sky over the
  course of time you can find out (almost) all there is to know about the
  universe.}
\end{quote}

\begin{acknowledgements}
  I thank Michael Perryman for his comments and Andreas Quirrenbach for
  reminding me that astrometric radial velocities are interesting in practice,
  not just in principle.

  The background image in figure \ref{fig:overview} was obtained from the
  Digitized Sky Surveys, which were produced at the Space Telescope Science
  Institute under U.S. Government grant NAG W-2166. The Second Palomar
  Observatory Sky Survey (POSS-II) was made by the California Institute of
  Technology with funds from the National Science Foundation, the National
  Geographic Society, the Sloan Foundation, the Samuel Oschin Foundation, and
  the Eastman Kodak Corporation. The Oschin Schmidt Telescope is operated by the
  California Institute of Technology and Palomar Observatory.
\end{acknowledgements}

\end{document}